\documentclass[aps,pra,showpacs]{revtex4}
\usepackage{graphicx}
\begin{document}
\def\la{{\langle}}
\def\ra{{\rangle}}
\def\vep{{\varepsilon}}
\newcommand{\beq}{\begin{equation}}
\newcommand{\eeq}{\end{equation}}
\newcommand{\beqa}{\begin{eqnarray}}
\newcommand{\eeqa}{\end{eqnarray}}
\newcommand{\da}{^\dagger}
\newcommand{\wh}{\widehat}

\title{On first-arrival-time distributions for a Dirac
electron in 1+1 dimensions}
\author{D. Alonso}
\affiliation{Departamento de F\'{\i}sica Fundamental
y Experimental, 
\\Electr\'onica y Sistemas. Universidad de La Laguna,
La Laguna 38203, Tenerife, Spain}
\author{R. Sala Mayato}
\affiliation{Departamento de F\'{\i}sica Fundamental II.
Universidad de La Laguna,
La Laguna 38203, Tenerife, Spain}
\author{C. R. Leavens}
\affiliation{Institute for Microstructural Sciences,
National Research Council of Canada, Ottawa, Canada
K1A 0R6}

\begin{abstract}
%
For the special case of freely evolving Dirac electrons in $1\;+\;1$
dimensions, Feynman checkerboard paths have previously been used to
derive Wigner's arrival-time distribution which includes all arrivals.
Here, an attempt is made to use these paths to determine the corresponding
distribution of first-arrival times. Simple analytic expressions are
obtained for the relevant components of the first-arrival propagator.
These are used to investigate the relative importance of the
first-arrival contribution
to the Wigner arrival-time distribution and of the contribution arising
from interference between first and later (i.e. second, third, ... ) arrivals. 
It is found that a distribution of (intrinsic) first-arrival
times for a Dirac electron cannot in general be consistently
defined using checkerboard paths, not even approximately in the
nonrelativistic regime.
\end{abstract}
\maketitle
%
%

\section{Introduction}

In the past decade there has been considerable interest in deriving and
understanding arrival-time distributions for quantum particles,
using a wide variety of approaches (for recent reviews see
\cite{rev1,rev2,MSE02}).
Here we focus on an approach based on Feynman paths \cite{feynman}.

Yamada and Takagi \cite{yam1,yam2} applied the consistent histories
approach,
with Feynman paths as particle histories, to the problem of deriving an
intrinsic arrival-time distribution. They considered the special case of
a
freely evolving nonrelativistic quantum particle in one spatial
dimension.
In \cite{yam1} they showed that
within their approach one cannot classify the histories according to the
number of times $n$ that a particle arrives at a given spatial point
$x=X$ during a specified finite time interval because the amplitude
for $n$ arrivals is zero for every finite $n$. Their qualitative explanation
was
that a typical Feynman path, being nondifferentiable in time, intersects
$X$ an infinite number of times in the given time interval, leaving zero
amplitude for any finite value of $n$. They further suggested that this
would
also be the case for a quantum particle propagating in three spatial
dimensions in the presence of a potential.

The velocity associated with a typical Feynman path for a
nonrelativistic
electron is infinite at almost every point on it. This is not the case
for a
relativistic electron -- the velocity associated with a Feynman
path \cite{feynman,JS} for a nonrelativistic electron is infinite at almost 
every point on it. Hence, such a path will not intersect $X$ an infinite
number of times in a given finite time interval and the amplitude for $n$
arrivals need not be zero for every finite $n$. This is the primary
motivation for the following attempt to derive the first-arrival-time
distribution for a freely
evolving Dirac electron in $1+1$ dimensions. In Sec II,
a checkerboard path derivation \cite{crl} of Wigner's arrival-time
distribution
\cite{Wigner} which includes all arrivals is sketched, primarily to
introduce
the basic notation and concepts used in the following sections.
In Sec. III the first-arrival propagator
for this special case is derived. Unfortunately, there is interference
between the $n=1$ and $n>1$ contributions to the arrival-time distribution
so that the distribution of intrinsic first-arrival times cannot be consistently
defined using this approach. Calculated results for the $n=1$ and
interference contributions are presented in Sec. IV 
for two simple cases. Concluding remarks are made in Sec. V.

\section{Checkerboard Paths and Wigner's arrival-time distribution for
Dirac electrons}

Consider the $1+1$ dimensional free-electron Dirac equation in the form
\beq\label{CB1}
i\hbar\hat{1}\frac{\partial}{\partial t}
\Psi(x,t)\:=\:\hat{H}\Psi(x,t)\:=\:
-i\hbar c \hat{\sigma}_z
\frac{\partial}{\partial x} \Psi(x,t)\:-\:mc^2\hat{\sigma}_x\Psi(x,t)
\eeq
with $\Psi(x,t)\:\equiv\:\bigl (\Psi_{+}(x,t),\Psi_{-}(x,t)\bigr )^T$ a
two-component spinor and
\beq
\hat{\sigma}_x\:=\:\left(\matrix{0&1\cr 1&0}\right)\;\;,\;\;
\hat{\sigma}_z\:=\:\left(\matrix{1&0\cr 0&-1}\right)\;\;.
\eeq

The velocity operator
\beq\label{VO}
\hat{v}\:\equiv\:\frac{d\hat{x}}{dt}\:=\:\frac{i}{\hbar}\:[\hat{H},\hat{x}]\;
=\:c\:\hat{\sigma}_z
\eeq
has the orthonormal eigenfunctions
\beq
\chi_{ _+}\:=\:\left(\matrix{1 \cr 0}\right)\;\;,\;\;
\chi_{ _-}\:=\:\left(\matrix{0 \cr 1}\right)
\eeq
with eigenvalues $+c$ and $-c$ respectively. Hence,
\beq
\Psi(x,t)\:=\:\left (\matrix{\Psi_{+}(x,t) \cr \Psi_{-}(x,t)}\right )
\:=\:\Psi_{+}(x,t)\;\chi_{+} + \Psi_{-}(x,t)\;\chi_{-}
\eeq
and the upper and lower components of the wave function $\Psi(x,t)$ are,
respectively, the amplitudes at ($x,t$) for the right-going
($+c$) and left-going ($-c$) velocity eigenstates $\chi_{+}$ and
$\chi_{-}$.
The four components of the retarded $2\times 2$ propagator
$K(x_B,t_B;x_A,t_A)\equiv K(B;A)$  of the $1+1$ dimensional
free-electron Dirac equation
(\ref{CB1}) are labelled by velocity directions ($+$ or $-$) at
$(x_A,t_A)$ and at $(x_B,t_B)$ and are accordingly defined by
$$\left(\matrix{\Psi_+(x_B,t_B)\cr \Psi_-(x_B,t_B)}\right)\:=\:
\int_{-\infty}^{\infty}dx_A\left(\matrix{K_{++}(x_B,t_B;x_A,t_A)&
K_{+-}(x_B,t_B;x_A,t_A)\cr
K_{-+}(x_B,t_B;x_A,t_A)&K_{--}(x_B,t_B;x_A,t_A)}
\right)\:\left(\matrix{\Psi_+(x_A,t_A)\cr \Psi_-(x_A,t_A)}\right)\:.
$$
The left subscript, $+$ or $-$, on $K$ denotes respectively a
right-going
or left-going arrival at $x_B$ at time $t_B$ while the right subscript,
$+$ or $-$, denotes respectively a
right-going or left-going departure from $x_A$ at time $t_A$.

Subtracting the Hermitean conjugate of (\ref{CB1}), multiplied from the
right by
$\Psi(x,t)$, from (\ref{CB1}) multiplied from the left by the Hermitean
conjugate of $\Psi(x,t)$ gives the continuity equation
\beq\label{CB3}
\frac{\partial}{\partial t} \Psi^{\dagger}(x,t)\Psi(x,t)\:+\:
\frac{\partial}{\partial x}
\Psi^{\dagger}(x,t)c\hat{\sigma}_z\Psi(x,t)\:=0\;.
\eeq
It is assumed throughout the paper that the parameters of the initial
wave function are such that Dirac's original identification of  
\beq\label{CB4}
\rho(x,t)\:=\:\Psi^{\dagger}(x,t)\Psi(x,t)\:=|\Psi_{+}(x,t)|^2\:+\:
|\Psi_{-}(x,t)|^2
\eeq
and
\beq\label{CB4J}
J(x,t)=c\Psi^{\dagger}(x,t)\:\hat{\sigma}_z\Psi(x,t)\;=\;c\:[|\Psi_{+}(x,t)|^2
\:-\:|\Psi_{-}(x,t)|^2]
\eeq
with single-electron probability and probability current densities,
respectively, is an adequate approximation.

In Feynman and Hibb's  classic book ``Quantum Mechanics and Path
Integrals''
\cite{feynman} it is stated that the free-electron
propagator $K(x_B,t_B;x_A,t_A)$ can be constructed from a model in
which a particle going from $x_A$ at time $t_A$ to
$x_B$ at time $t_B$ is constrained to move diagonally in space-time at
constant speed $c$ in checker fashion (i.e., forward in time with spatial
increment $\pm \Delta x$ with $\Delta x = c \Delta t= c (t_B-t_A)/N$ for
each of $N$ equal time steps $\Delta t >0$)\footnote
{In the following, the word ``particle'' is  reserved for the
mathematical entity of the checkerboard model to distinguish it
from the physical entity that is being timed and
which is referred to by the words ``electron'' or ``quantum particle''.}.
Each component of the propagator is obtained as the
$N\:\rightarrow \:\infty$ limit of the sum over all $N$-step
checkerboard
paths joining ($x_A,t_A$) to ($x_B,t_B$), with the first and last steps
appropriately
fixed, when the weight associated with a path having $R$ (noncompulsary)
reversals of direction or corners is taken to be $(imc^2\Delta t/\hbar)^R\;$.
Jacobson and Schulman \cite{JS} regrouped the sum-over-paths
into a sum-over-R, i.e.
\beq\label{KJS}
K_{\beta \alpha}(B;A)= i\:(mc/2\hbar)\:\lim_{N \to \infty}
\:\sum_{R \ge 0}
\Phi_{\beta \alpha}(R) (imc^2 \Delta t/\hbar)^R \;\;
(\alpha = \pm ; \beta = \pm)\;,
\end{equation}
where $\Phi_{\beta \alpha}(R)$ is the number of $\beta \alpha$ paths with
$R$ noncompulsary reversals.
They also evaluated the four checkerboard path integrals, obtaining the
following closed-form expressions for the components of the propagator:
\beq\label{KRR}
K_{++} (B;A)\:=-\:\frac{c\:t_{BA}+x_{BA}}{2\lambda_c\;
l_{BA}}\:J_1\biggl(\frac{l_{BA}}{\lambda_c}\biggr)\;,
\eeq
\beq\label{KLL}
K_{--}(B;A)\:=-\:\frac{c\:t_{BA}-x_{BA}}{2\lambda_c\;
l_{BA}}\:
J_1\biggl ( \frac{l_{BA}}{\lambda_c}\biggr )\;,
\eeq
\beq\label{KRL}
K_{+-}(B;A)\:=\:K_{-+}(B;A)\:=\frac{i}{2\lambda_c}\:
J_0 \biggl (\frac{l_{BA}}{\lambda_c} \biggr )\; ,
\eeq
where $\lambda_c \equiv \hbar/mc$ is the Compton wavelength of the electron,
$x_{BA}\equiv (x_B - x_A)$, $t_{BA}\equiv (t_B-t_A)$ and
$l_{BA} \equiv c \tau_{BA}$ with
$\tau_{BA} \equiv t_{BA}[1 -(v_{BA}/c)^2]^{1/2}$
the proper time for a particle moving with constant velocity
$v_{BA}\equiv x_{BA}/t_{BA}$.
Jacobson and Schulman also determined the number of reversals, $R_0$, that
gives the maximum contribution to the sum in (\ref{KJS}). They obtained
$R_0 = l_{BA} /\lambda_c $ and also showed
that the sum is dominated by terms
having $R$ within $\approx \!R_0^{1/2}$
of $R_0$ \footnote{It has apparently been
assumed that $|x_{BA}|\:<<\:ct_{BA}$ so that $R_0\:>>\:1$.}. 
The picture that emerges \cite{JS,GJKS,SPSAG} is one in which the particle
always moves with speed $c$ and typically travels a distance $\approx \!
(t_{BA}/\tau_{BA})\lambda_c \geq \lambda_c$
between reversals of direction; its motion
is Brownian with diffusion constant $\hbar/m$ only on scales much larger
than this correlation distance.

Now consider the problem of deriving an expression for the distribution
$\Pi(T;X)$ of arrival times $T$ at the spatial point $x=X$ for an ensemble of
Dirac electrons all prepared in the same initial state $\Psi(x,0)$. Following
Yamada and Takagi, it is {\it assumed} that the arrival-time distribution
for the fictitious particles of the checkerboard model (with $t_A=0$,
$x_B=X$ and $t_B=T$), should it be well-defined, can be identified with the
desired distribution for actual electrons.  It should be noted
that the expression (\ref{CB4}) for $\rho(x,t)$ contains
no $+/-$ cross terms arising from interference between paths
arriving at $(x,t)$ with right-going ($+$) and those with left-going
($-$)
velocities. Hence, at least for the particles of the
checkerboard model, the probability
density $\rho(x,t)$ can be decomposed into two contributions, one
associated
only with right-going arrivals at $(x,t)$ and the
other only with left-going arrivals:
$\rho(x,t)\:=\:\rho_{+}(x,t)\:+\:\rho_{-}(x,t)$ with
$\rho_{\pm}(x,t)\:\equiv\:|\Psi_{\pm}(x,t)|^2\;$. Now, recall that particles
following checkerboard paths move only at speed $c$ and that the time
between reversals in their directions of motion is
$\approx \!(t_{BA}/\tau_{BA}) \lambda_c/c$ for those paths
that make the dominant contribution to the propagator. Hence, for
$\Delta t$
much less than this correlation time, nearly all of the particles
in the spatial interval $[X-c\Delta t,X]$ that are right-going
at time $T-\Delta t$ should arrive
at $x=X$ during the time interval $[T-\Delta t, T]$, giving the dominant
contribution to the number of right-going arrivals at $X$ during that
time
interval. Taking the limit $\Delta t \rightarrow dt = 0^+$ this leads to
the
prediction that $c\:\rho_{+}(X,T)\:dt$
right-going particles arrive at $x=X$ during the infinitesimal
time interval $[T-dt,T]$. A similar argument applies for left-going
particles
in $[X, X+c\Delta t]$ at time $T-\Delta t$. Hence, the distribution of
arrival times $T$ at the spatial point $x=X$ is given by

\beq\label{cb12}
\Pi(T;X)\:=\:\Pi_{+}(T;X)\:+\:\Pi_{-}(T;X)\; ; \;
\Pi_{\pm}(T;X)\:=\:|\Psi_{\pm}(X,T)|^2 \; \bigg / \; \int_0^{T_{max}}dt \rho(X,t)\;,
\eeq
where $T_{max}$ is the maximum arrival time of practical interest.
For the special case of free motion in one spatial dimension, the general
results presented without derivation or discussion by Wigner \cite{Wigner}
simplify to (\ref{cb12}) \cite{crl}.

\section{First-arrival propagator for a Dirac particle in 1+1
dimensions}

The fundamental constants $\hbar$ and $c$ are set to $1$ in the analysis of
this section but restored in the final results.

\begin{figure}
{\includegraphics[width=3.35in]{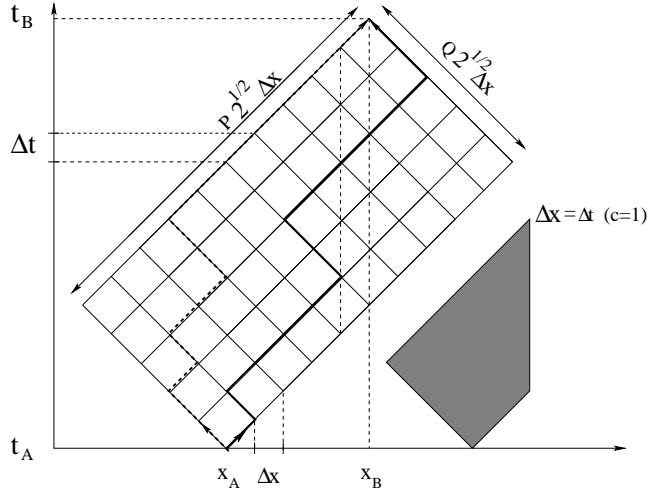}}
\caption[]{Checkerboard grid with $\Delta x = \Delta t$ in the $x-t$
plane $(c=1)$. Two paths with five corners, the last of which is compulsary,
are shown.
The solid-line path with $\alpha=+$ and $\beta=-$ does not contribute to
the first-arrival propagator $K^{(1)}(x_B,t_B;x_A,t_A)$ while the
dashed-line path with $\alpha=-$ and $\beta=+$ does contribute.
The restricted domain for evaluation of the first-arrival propagator is
bounded on the right by the vertical dotted line; its shape is shown in grey.}
\protect\label{pathgrid2}
\end{figure}

To begin, suppose that the spatial interval of interest is divided into
$M$ pieces each of length $\Delta x=x_{BA}/M$, assuming $x_B > x_A$.
Suppose further that a particular path of $N$ time-steps
consists of $P$ spatial steps of length $\Delta x$ to the right and $Q$
to the left so that $N=P+Q$ and $M=P-Q$. The resulting space-time grid of
path segments available to the particle is illustrated in
Fig. \ref{pathgrid2}. Denote by $K_{\beta\alpha}^{(1)}(B;A)$ the
component of the propagator (\ref{KJS}) associated with first arrivals
at $X=x_B$ at time $T=t_B$. It is constructed from only those
paths that reach $x_B$ for the {\it{first time}} at time $t_B$.
To compute $K_{\beta \alpha}^{(1)}$ we have to count the number of
$\beta \alpha$ paths with
$R$ noncompulsary reversals for the restricted space-time grid shown in Fig.
\ref{pathgrid2}.
For simplicity it is assumed that the initial wave function
$\Psi(x,t_A=0)$
is sufficiently well localized to the left of $x_B=X$ that
$K_{++}^{(1)}(B;A)$ and $K_{+-}^{(1)}(B;A)$ are the
only components of the first-arrival propagator that need to be considered.
For $x_A < x_B$ the components $K_{-+}(B;A)$ and $K_{--}(B;A)$ of
course have contributions only from multiple-arrival paths.

\subsection{Counting the number of first arrival paths with $R$ corners}

\begin{figure}
{\includegraphics[width=3.35in]{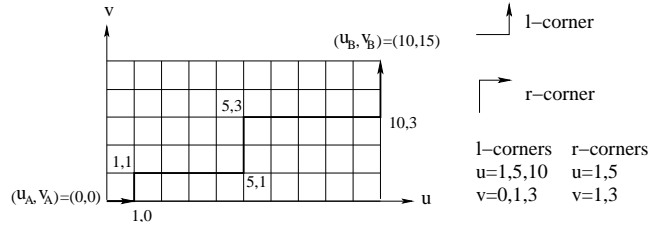}}
\caption[]{Example of an unrestricted path that would not be allowed once
first-arrival restrictions are introduced. The $u$ and $v$
sequences corresponding to the three $l$-corners and two $r$-corners
 of the path are indicated.}
\label{fullgrid}
\end{figure}

Computation of $\Phi_{\beta \alpha}^{(1)}(R)$ involves counting the
number of restricted $\beta\alpha$ paths with a given number of
corners in a lattice. A path in
the lattice of Fig. \ref{fullgrid} -- obtained from
Fig. \ref{pathgrid2} by clockwise rotation through $45^{ o}$ and
rescaling by a factor of $(2^{1/2} \Delta x)^{-1}$ --  is built up of
$(1,0)=\; \rightarrow$ and $(0,1)=\; \uparrow$ motions. There
are no $\leftarrow$ and $\downarrow$ motions because the paths of Fig. \ref{pathgrid2}
move only forward in time. Denote by a $l$-corner a point on the path
that is reached by a $\rightarrow$ step and is left by a $\uparrow$ step
and by a $r$-corner a point that is reached by a $\uparrow$ step
and left by a $\rightarrow$ step. Denote by $R_l$
the number of corners of type $l$ and by $R_r$ the number of
corners of type $r$ in a given path. Any  path in the $u-v$ lattice with given initial
and final points $(u_A,v_A)$ and $(u_B,v_B)$ can be completely specified by either the
coordinates of its $l$ corners or by the coordinates of its $r$ corners. Both specifications
are needed in the following derivation of the first-arrival propagator.

First, consider the counting problem
without any $\beta \alpha$ or first-arrival restrictions on the paths
and including compulsary as well as noncompulsary corners.
In this simple case, enumeration of the number of paths in a lattice with a
given number of $l$- or $r$-corners is solved in the following
manner \cite{kra}. To count the number of paths with $R_l$ $l$-corners one
builds two vectors ${\bf{u}}$ and ${\bf{v}}$ that contain the integer $u$ and
$v$-coordinates of the $R_l$ corners of such a path:
\begin{eqnarray}
{\bf{u}}=(u_1,u_2,\cdots,u_{R_l}) \cr
{\bf{v}}=(v_1,v_2,\cdots,v_{R_l}).
\end{eqnarray}
These coordinates satisify the inequalities
\begin{eqnarray}
\label{ine1}
u_A +1  \le u_1<u_2 \cdots <u_{R_l} \le u_B \cr
v_A \le v_1<v_2 \cdots <v_{R_l} \le v_B - 1,
\end{eqnarray}
where the sequences are strictly ordered. Denote by $\Phi_l^{wr}(R_l)$ the
number of paths with $R_l$ $l$-corners, where the label $"wr"$ is a
reminder that the paths are without
restriction. This number may be evaluated by observing
that there are $u_B-u_A$ integers from which to choose the $u$-coordinates
and $v_B-v_A$ from which to choose the $v$-coordinates. The required number is then
\begin{equation}
\label{nores}
\Phi_l^{wr}(R_l)=\pmatrix{u_B-u_A \cr R_l} \pmatrix{v_B-v_A \cr R_l}.
\end{equation}
Similarly, one obtains
\begin{equation}
\label{noresp}
\Phi_r^{wr}(R_r)=\pmatrix{u_B-u_A \cr R_r} \pmatrix{v_B-v_A \cr R_r}.
\end{equation}

\begin{figure}
{\includegraphics[width=3.35in]{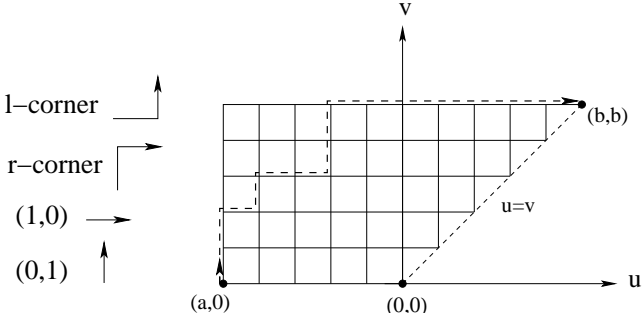}}
\caption[]{A restricted path from $(u_A,v_A)=(a,0)$ to $(u_B,v_B)=(b,b)$
which has two $l$-corners and three $r$-corners.}
\label{pathcount}
\end{figure}

Now consider only those paths which do not {\it cross} (touching is allowed
for the moment)
$x=x_B$ before $t=t_B$. When considering such restricted paths in the $u-v$
lattice (see Fig. \ref{pathcount}) it is convenient to choose the bottom-right
corner of the accessible region as the origin of the $u-v$ coordinate system
so that the region below the diagonal $u=v$ is out-of-bounds.

First consider the case in which the paths are specified by the coordinates of their
$l$-corners. This case is the easier of the two because $l$-corners are diagnostic, i.e.
a necessary and sufficient condition for a restricted path is that none of its $l$-corners
be on the forbidden $u>v$ side of the diagonal.  It is thus
required to calculate the number of paths with $R_l$ $l$-corners from
$(u_A,v_A)$ to $(u_B,v_B)$ with $u_A=a$, $v_A=0$ and $u_B=v_B=b$, say,
such that $v_i\ge u_i$ for all $i$. This number, $\Phi_l(R_l;BA)$, is the number (\ref{nores})
minus the number of paths with $R_l$ $l$-corners such that $u_i > v_i$ for at least one $i$.
To calculate the latter number, take $k$ to be the largest integer such that
$u_k > v_k$ and build from (\ref{ine1}) the following sequences
\begin{eqnarray}
\label{ine2}
u_A + 1&=&a+1\le u_1<u_2 \cdots <u_{k-1} <v_{k+1} \cdots <v_{R_l}\le v_B-1
\;=\;b-1\cr
v_A&=&0\le v_1<v_2 \cdots <v_k < u_k< u_{k+1} \cdots <u_{R_l}\le u_B \;= \;b\;.
\end{eqnarray}
The sequences are ordered (as can be checked, see \cite{kra}) and there is
a one-to-one correspondence between the double sequences (\ref{ine1}) and
the double sequences (\ref{ine2}). The number of all the double sequences
(\ref{ine2}) is
\begin{equation}
\pmatrix{b-a-1 \cr R_l-1} \pmatrix{b+1\cr R_l+1},
\end{equation}
and therefore $\Phi_l(R_l;BA)$ is given by
\begin{equation}\label{phil}
\Phi_l(R_l;BA)=\pmatrix{b-a \cr R_l} \pmatrix{b \cr
R_l}-\pmatrix{b-a-1 \cr R_l-1} \pmatrix{b +1\cr R_l+1}.
\end{equation}

Now consider, the computation of the corresponding number of restricted
paths from $(a,0)$ to $(b,b)$ with $R_r$ $r$-corners, i.e. $\Phi_r(R_r;BA)$.
A necessary condition for such a path is, of course, that all of its
$R_r$  $r$-corners be on the allowed $v>u$ side of the diagonal. However, this is not a 
sufficient condition because it is possible for the $l$-corner between two
consecutive such $r$-corners to be on the forbidden side of the diagonal. An additional
complication is that the first $l$-corner might precede the first  $r$-corner and/or the 
last $l$-corner might follow the last $r$-corner. A simple way to include such $l$-corners in
the analysis is to add the end-points $(a,0)\equiv (u_0,v_0)$ and $(b,b)
\equiv (u_{R_r +1},v_{R_r + 1})$ to the set of $R_r$ $r$-corners to obtain the set of $R_r +2$
points $\{ (u_0,v_0),(u_1,v_1), ... , (u_{R_r},v_{R_r}),(u_{R_r +1},v_{R_r +1})\}$. Now,
for $1\leq i\leq R_r - 1$, $(u_{i+1},v_i)$ is the (diagnostic) $l$-corner between the
consecutive r-corners $(u_i,v_i)$ and $(u_{i+1},v_{i+1})$. Hence,  
$\Phi_r(R_r;BA)$ is the number (\ref{noresp}) minus the
number of paths with $R_r$ $r$-corners such that $u_{i+1} > v_i$ for at least one $i$, with
$0\leq i \leq R_r$ to allow for paths for which the first and/or last corner is an $l$-corner.
To calculate the latter number, take $k$ to be the smallest integer such
that $u_{k+1} > v_k$ and construct the sequences
\begin{eqnarray}
a=u_0\leq u_1 < u_2 < ... < u_k < v_{k+1} < ... <v_{R_r}\leq v_{R_r + 1} = b  \cr
1=v_0 +1 \leq v_1 < v_2 < ... <v_k < u_{k+1} < ... < u_{R_r} \leq u_{R_r +1} -1 = b-1 \;.
\end{eqnarray}
The total number of these double sequences is
\begin{equation}
\pmatrix{b-a+1\cr R_r} \pmatrix{b-1 \cr R_r}\;,
\end{equation}
and therefore $\Phi_r(R_r;BA)$ is given by
\begin{equation}\label{phir}
\Phi_r(R_r;BA)=\pmatrix{b-a \cr R_r} \pmatrix{b \cr R_r}-
\pmatrix{b-a +1\cr R_r} \pmatrix{b - 1\cr R_r}\;.
\end{equation}

Finally, consider the desired $\beta \alpha$ first-arrival paths, which
are not allowed to {\it touch} $x=x_B$ before $t=t_B$. The various
notations $A$, $(x_A,t_A)$ and $(u_A,v_A)2^{1/2}\Delta x$ are reserved for the first point
of a checkerboard path and $B$, $(x_B,t_B)$ and $(u_B,v_B)2^{1/2}\Delta x$ for the last point.
Denote by $A_{\alpha}$, $(x_A +\alpha \Delta x, t_A+\Delta t)$ and
$(a_{\alpha},0)2^{1/2}\Delta x$ the point on a $\beta\alpha$ path at time $t=t_A + \Delta t$
and by $B_{\beta}$, $(x_B-\beta \Delta x, t_B-\Delta t)$,
$(b_{\beta},b_{\beta})2^{1/2} \Delta x$ the point
at time $t=t_B - \Delta t$.

It is important to note that those paths which may touch but do not cross the
diagonal $u=v$, extending from $(0,0)$ to $(b_{\beta},b_{\beta})$, do not
touch $x=x_B$ before $t=t_B$ and hence are first-arrival paths. Hence,
the above expressions for $\Phi_l(R_l;BA)$ and $\Phi_r(R_r;BA)$ with $A$, $B$,
$a$ and $b$ replaced by $A_{\alpha}$, $B_{\beta}$, $a_{\alpha}$ and
$b_{\beta}$, respectively, can be used to evaluate
$\Phi^{(1)}_{\beta \alpha}$ and hence $K^{(1)}_{\beta \alpha}$. It is
only necessary, for a given choice of $\alpha$ and $\beta$, to determine
$a_{\alpha}$, $b_{\beta}$ and the relation between
$R$ and $R_l$ or $R_r$, keeping in mind that $R$ includes only noncompulsary
corners.

\begin{figure}
{\includegraphics[width=3.35in]{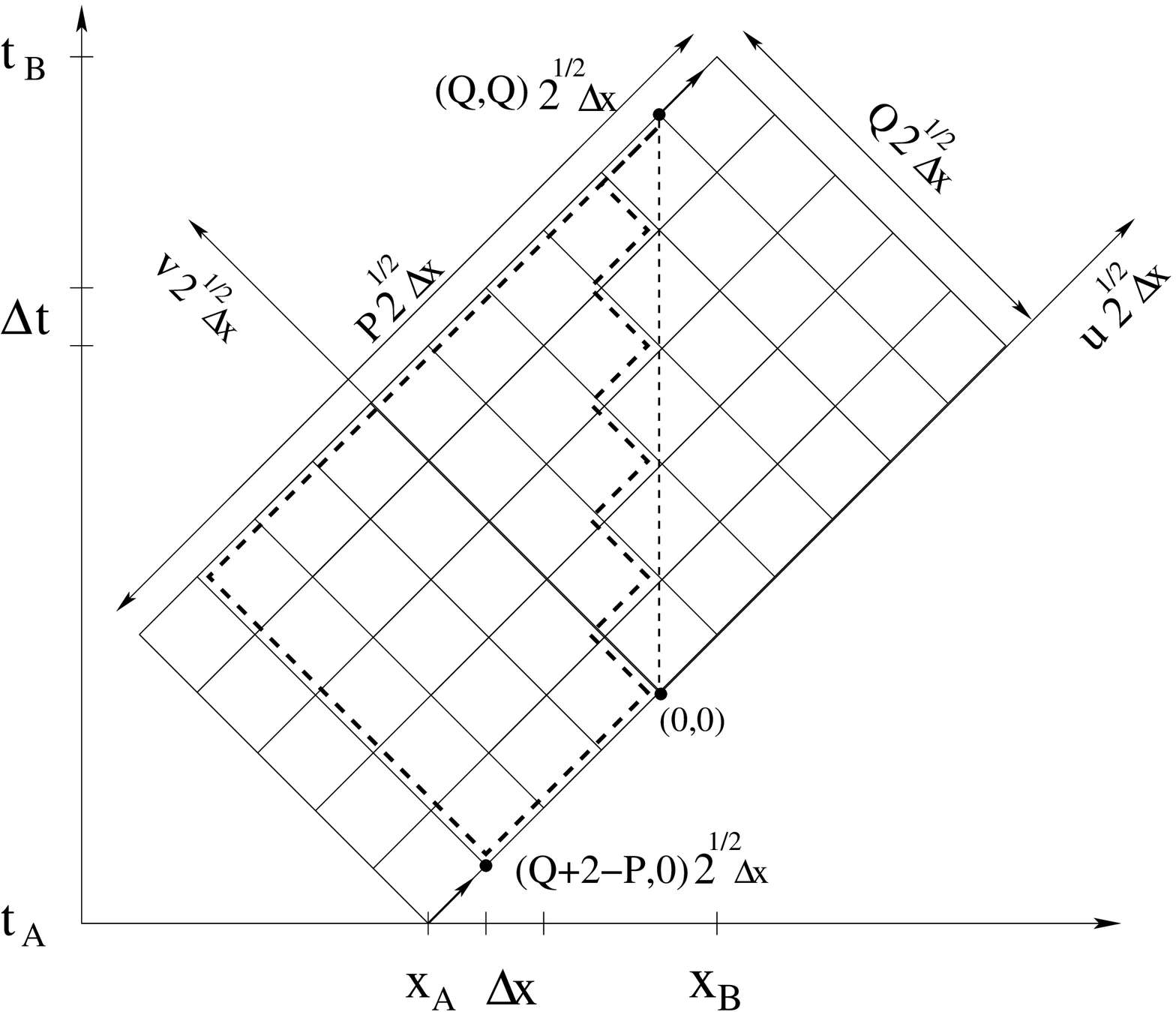}}
\caption[]{The dashed line demarcates the grid to be considered in the
enumeration problem related to $K_{++}^{(1)}$.}
\label{pathgridpp}
\end{figure}

For $\beta \alpha\: =\: ++$ it is clear from Fig. \ref{pathgridpp}
that $(b_+,b_+)=(Q,Q)$, $(u_B,v_B)=(Q+1,Q)$, $(u_A,v_A)=(Q+1-P,0)$ and
$(a_+,0)=(Q+2-P,0)$. Then a little thought leads to the conclusion
that the number $\Phi_{++}^{(1)}(R)$ is equal to $\Phi_r(R_r;B_+A_+)$
with the identification $R=2R_r -1$. Hence, upon replacing $a$ by $a_+=Q+2-P$,
$b$ by $b_+=Q$, and $R_r$ by $(R+1)/2$ in (\ref{phir}), it follows that
\begin{eqnarray}
\Phi_{++}^{(1)}(R)&=&\pmatrix{P-2 \cr (R+1)/2} \pmatrix{Q \cr (R+1)/2}-
\pmatrix{P-1 \cr (R+1)/2} \pmatrix{Q-1 \cr (R+1)/2} \cr \cr
&=&\pmatrix{P-1 \cr (R+1)/2} \pmatrix{Q-1 \cr (R-1)/2}-
\pmatrix{P-2 \cr (R-1)/2} \pmatrix{Q \cr (R+1)/2},
\end{eqnarray}
where the identity $    \pmatrix{n-1 \cr k}= \pmatrix{n \cr k}-\pmatrix{n-1
\cr k-1}$ has been used.

\begin{figure}
{\includegraphics[width=3.35in]{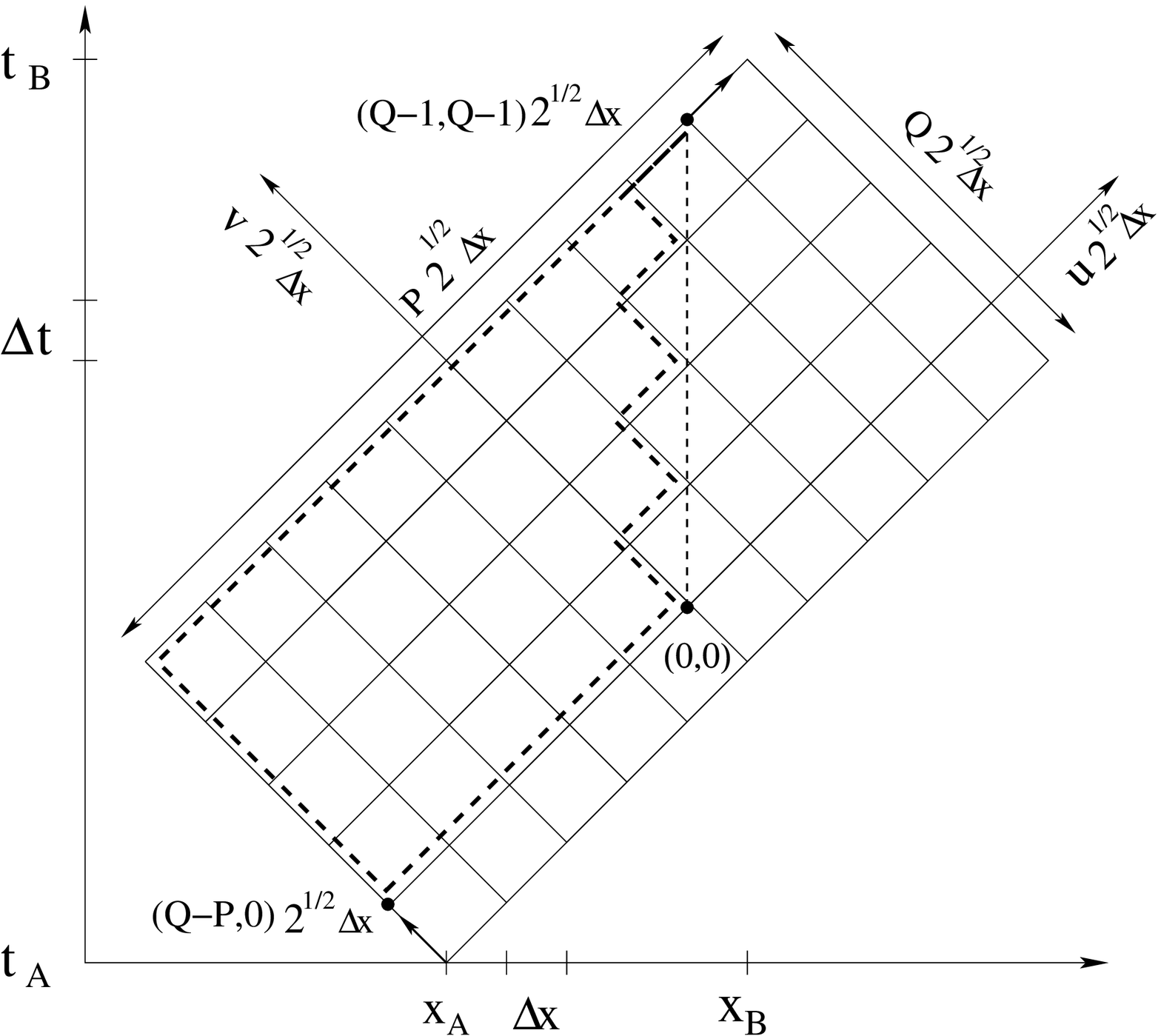}}
\caption[]{The dashed line demarcates the grid to be considered in the
enumeration problem related to $K_{+-}^{(1)}$.}
\label{pathgridpm}
\end{figure}

For $\beta \alpha\: =\:+-$ it is clear
from Fig. \ref{pathgridpm} that $(b_+,b_+)= (Q-1,Q-1)$,
$(u_B,v_B)=(Q,Q-1)$, $(a_-,0)=(Q-P,0)$, $(u_A,v_A)=(Q-P,-1)$ and
$\Phi_{+-}^{(1)}(R)$ is equal to $\Phi_l(R_l;B_+ A_-)$ with $R=2R_l$.
Hence, upon replacing $a$ by $a_- = Q-P$, $b$ by $b_+=Q-1$ and $R_l$ by
$R/2$ in (\ref{phil}), it follows that
\begin{equation}
\Phi_{+-}^{(1)}(R)=\pmatrix{P-1 \cr R/2} \pmatrix{Q-1 \cr R/2}-
\pmatrix{P-2 \cr (R-2)/2} \pmatrix{Q \cr (R+2)/2},
\end{equation}
with $\Phi_{+-}^{(1)}(1)=1$.

\subsection{Evaluation of the propagators}

The first-arrival-time propagators are expressed
as
\begin{eqnarray}
\label{prop1st}
K_{++}^{(1)}(B;A)=\frac{i}{2\lambda_c}\: \lim_{N \to \infty}
\sum_{odd\; R \ge 1}\Phi_{++}^{(1)}(R) (i m \Delta t)^R \cr
K_{+-}^{(1)}(B;A)= \frac{i}{2\lambda_c}\:\lim_{N \to \infty}
\sum_{even\; R \ge 0}\Phi_{+-}^{(1)}(R) (i m \Delta t)^R.
\end{eqnarray}

\subsection{The case of $K_{++}^{(1)}$}

In this case we have from (\ref{prop1st}) and using the approximation
$\pmatrix{n \cr m} \approx n^m/m!$, that becomes exact as $n \to \infty$,
\begin{eqnarray}
K_{++}^{(1)}(B;A)&=&\frac{i}{2\lambda_c} \lim_{N \to \infty}
\sum_{odd\;R\:\ge 1}(i m \Delta t)^R \Bigg( \frac{P^{(R+1)/2}}{[(R+1)/2]!}
\frac{Q^{(R-1)/2}}{[(R-1)/2]!}-\frac{P^{(R-1)/2}}{[(R-1)/2]!}
\frac{Q^{(R+1)/2}}{[(R+1)/2]!}
\Bigg) \cr
&=& \frac{i}{2\lambda_c}\lim_{N \to \infty}
\sum_{odd\;R\ge 1}(im \Delta t)^R (P-Q)\frac{(PQ)^{(R-1)/2}}
{[(R+1)/2]! ([(R-1)/2]!}.
\end{eqnarray}
This expression may be transformed \footnote{From $P+Q\equiv N \equiv
t_{BA}/\Delta t$ and $P-Q\equiv M \equiv x_{BA}/c\Delta t$ it readily follows
that $M/N=x_{BA}/ct_{BA}\equiv v_{BA}/c$ and $(c\Delta t)^2 4PQ=
(c\Delta t)^2(N^2 - M^2)=(ct_{BA})^2[1-(v_{BA}/c)^2]\equiv l_{BA}^{\;2}$.} to
\begin{eqnarray}
\label{propp}
K_{++}^{(1)}(B;A)&=& i \frac{x_{BA}}{\lambda_c l_{BA}}\:
\lim_{N \to \infty}
\sum_{odd\:R\ge 1} \biggl(\frac{il_{BA}}{2 \lambda_c}\biggr)^R
\frac{1}{[(R+1)/2]!
[(R-1)/2]!} \cr
&=& - \frac{x_{BA}}{\lambda_c l_{BA}}\:\cdot\:\frac{l_{BA}}{2\lambda_c}
\sum_{k=0}^{\infty} (-)^k
\biggl(\frac{l_{BA}}{2 \lambda_c}\biggr)^{2k}
\frac{1}{k!(k+1)!}\;.
\end{eqnarray}
In the last line $R$ has been replaced by $2k+1$ and the limit $N
\rightarrow \infty$ taken. Comparison with the power series representation
of the Bessel function,
\begin{equation}\label{sum}
J_n(z)=(\frac{z}{2})^n\sum_{k=0}^{\infty}\frac{(-)^k}{k! (k+n)!}
(\frac{z}{2})^{2k},
\end{equation}
immediately gives
\begin{equation}\label{++1}
K_{++}^{(1)}(B;A)=-\frac{x_{BA}}{\lambda_c l_{BA}}
J_1 \biggl(\frac{l_{BA}}{\lambda_c}\biggr)\;.
\end{equation}

\subsection{The case of $K_{+-}^{(1)}$}

In this case we start with the expression
\begin{eqnarray}
K_{+-}^{(1)}(B;A)= \frac{i}{2\lambda_c}\lim_{N \to \infty}
\sum_{even\:R\:\ge 0}(i m\Delta t)^R \Biggl( \frac{P^{R/2}}{(R/2)!}
\frac{Q^{R/2}}{(R/2)!}-\frac{P^{(R-2)/2}}{[(R-2)/2]!}
\frac{Q^{(R+2)/2}}{[(R+2)/2]!}
\Biggr).
\end{eqnarray}
Steps analogous to those above lead to
\begin{equation}\label{+-1}
K_{+-}^{(1)}(B;A)=\frac{i}{2\lambda_c}\; \Biggl[
J_0\biggl(\frac{l_{BA}}{\lambda_c}\biggr)+\frac{c-v_{BA}}{c+v_{BA}}\;
J_2 \biggl(\frac{l_{BA}}{\lambda_c}\biggr)\Biggr]\:.
\end{equation}

An interesting equality emerges from the above results, namely
\begin{equation}
K_{++}(B;A)\;-\;K_{--}(B;A)=K_{++}^{(1)}(B;A).
\end{equation}
In fact all paths contributing to the $K_{--}(B;A)$ component of the
propagator touch the line $x=x_B$ at least twice and therefore this set of
paths is complementary to the one contributing to $K_{++}^{(1)}(B;A)$ in the
limit of $N \to \infty$.

Finally,
\begin{eqnarray}
K_{++}^{(2,3,\cdots)}(B;A)=K_{++}(B;A)
-K_{++}^{(1)}(B;A)=K_{--}(B;A)&=&\frac{x_{BA}-ct_{BA}}
{2\lambda_c l_{BA}}\;J_1\biggl(\frac{l_{BA}}{\lambda_c}\biggr)
\cr
K_{+-}^{(2,3,\cdots)}(B;A)=K_{+-}(B;A)-
K_{+-}^{(1)}(B;A)&=&-\:
\frac{i}{2\lambda_c}\;
\frac{c-v_{BA}}{c+v_{BA}}\; J_2\biggl(\frac{l_{BA}}{\lambda_c}\biggr).
\end{eqnarray}

\section{Interference between first and later arrivals}

The decomposition $K_{\beta\alpha}(B;A)\:=\:K_{\beta\alpha}^{(1)}(B;A)\:+\:
K_{\beta\alpha}^{(2,3,...)}(B;A)$ according to first and later (second, third,
etc.) arrivals of a particle at
$x_B$ at $t_B$ leads immediately to the corresponding decomposition
$\Psi_{\beta}(x_B,t_B)\:=\:\Psi_{\beta}^{(1)}(x_B,t_B)\:+\:
\Psi_{\beta}^{(2,3,...)}(x_B,t_B)$ for the $\beta=\pm$ components of the
wave function $\Psi(x_B,t_B)$. Substitution of the latter expression,
with $t_A =0$, $x_B=X$ and $t_B=T$, into the result (\ref{cb12}) for
the arrival-time distribution gives
\begin{equation}\label{decom}
\Pi(T;X)\:=\:\Pi^{(1)}(T;X)\:+\:\Pi^{(2,3,...)}(T;X)\:+\:
\Pi^{(1\:\times\:2,3,...)}(T;X)\; .
\end{equation}
For the special case considered here in which the initial probability
density $\rho(x_A,0)$ is negligible for $x_A\geq X$,
\begin{equation}\label{one}
\Pi^{(1)}(T;X)=C\biggl|\int_{-\infty}^{X}\!dx_A\bigl[
K^{(1)}_{++}(X,T;x_A,0)\Psi_+(x_A,0)+K^{(1)}_{+-}(X,T;x_A,0)\Psi_-(x_A,0)
\bigr]\biggr|^2
\end{equation}
is the contribution of first arrivals,
\begin{eqnarray}\label{twothree}
\Pi^{(2,3,...)}(T;X)\!\!&=&\!\!C\biggl|\int_{-\infty}^{X}\!dx_A\bigl[
K^{(2,3,...)}_{++}(X,T;x_A,0)\Psi_+(x_A,0)+K^{(2,3,...)}_{+-}(X,T;x_A,0)
\Psi_-(x_A,0)\bigr]\biggr|^2 \cr
&+&\!\!C\biggl|\int_{-\infty}^{X}\!dx_A\bigl[K^{(2,3,...)}_{-+}(X,T;x_A,0)
\Psi_+(x_A,0)+K^{(2,3,...)}_{--}(X,T;x_A,0)\Psi_-(x_A,0)\bigr]
\biggr|^2 \cr
& &
\end{eqnarray}
is the contribution of later-arrivals and
\begin{eqnarray}\label{cross}
\Pi^{(1\:\times\:2,3,...)}(T;X)\!&=&\!2C\:\Re\biggl\{
\int_{-\infty}^{X}\!dx_A\bigl[K^{(1)}_{++}(X,T;x_A,0)
\Psi_+(x_A,0)+K^{(1)}_{+-}(X,T;x_A,0)\Psi_-(x_A,0)\bigr]^{*} \cr
&\times& \!\int_{-\infty}^{X}\!dx_A\bigl[
K^{(2,3,...)}_{++}(X,T;x_A,0)\Psi_+(x_A,0)+K^{(2.3,...)}_{+-}(X,T;x_A,0)
\Psi_-(x_A,0)\bigr]\biggr\} \cr
& &
\end{eqnarray}
is the contribution due to interference between first and later arrivals.
($C$ is the normalization factor appearing in (\ref{cb12})). Of particular
interest here is the magnitude of the interference contribution relative to
the first-arrival contribution in the regime $v_{BA}<<c$.

First, however, briefly consider the regime in which $v_{BA}$ is so
close to $c$ that the correlation distance $\lambda_c[1-(v_{BA}/c)^2]^{-1/2}$
for reversal of direction is sufficiently large that for a typical
checkerboard path there is insufficient time for more than one arrival at
$x_B$. To be more quantitative, assume that the initial amplitude
$\Psi_-(x_A,t_A)$ of the $-c$ velocity eigenstate is completely negligible
with respect to the initial amplitude $\Psi_+(x_A,t_A)$ of the $+c$ eigenstate
so that one need consider only $K_{++}(B;A)$ and $K_{-+}(B;A)$. Also, for
$(x_B,t_B)=(X,T)$ and $(x_A,t_A)=(x_A,0)$ assume that $x_{BA}=X-x_A$ is very
close to $ct_{BA}=cT$ for those values of $T$ for which $\Pi(T;X)$ is
nonnegligible and for those values of $x_A$ for which $\rho(x_A,0)$ is
nonnegligible. In this regime,
$K_{++}^{(1)}(B;A)/K_{++}(B;A)\approx 1-\delta/2$ and
$K_{++}^{(2,3,...)}(B;A)/K_{++}(B;A)\approx \delta/2$ where $v_{BA} \equiv
(1-\delta)c$ with $\delta << 1$. In addition, $|K_{-+}(B;A)/K_{++}(B;A)|
\approx(l_{BA}/2ct_{BA})J_0(l_{BA}/\lambda_c)/J_1(l_{BA}/\lambda_c)$ with
$l_{BA}\approx (2\delta)^{1/2}ct_{BA}$. If $\delta$ is sufficiently small that
$l_{BA} << \lambda_c$ then, using the leading term in (\ref{sum}) for $n=0$
and for $n=1$, $|K_{-+}(B;A)/K_{++}(B;A)|\approx \lambda_c/x_{BA}
=\lambda_c/(X-x_A)$ which is typically very much less than $1$ for an initial
wave packet $\Psi(x_A,0)$ that is well-localized away from $x=X$.
Hence, at least to the extent that the concepts of  single-particle
probability and probability current densities are still meaningful in the
regime in which $v_{BA}$ is very close to $c$, the interference term is
very small and $\Pi(T;X)=\Pi^{(1)}(T;X)$ to a good approximation. Strictly
speaking, however, for the special case under consideration the
first-arrival-time distribution is well-defined only in the limit
$v_{BA}\rightarrow c$.

Now, consider the nonrelativistic regime. With the definitions $\Psi_{\pm}(x,t)\equiv
\phi_{\pm}(x,t)\exp(-imc^2t/\hbar)$, $\phi(x,t)\equiv \phi_+(x,t)+\phi_-(x,t)$ and
$\Delta \phi(x,t)\equiv \phi_+(x,t)-\phi_-(x,t)$, the $1$ + $1$ dimensional free-electron
Dirac equation (\ref{CB1}) can be written 
\begin{eqnarray}\label{nonrel}
i\hbar\:\partial \phi(x,t)/\partial t\:&=&\:-i\hbar c\:\partial \Delta \phi(x,t)/\partial x
-2mc^2 \phi(x,t)\; , \cr
i\hbar\:\partial \Delta \phi(x,t)/\partial t\:&=&\:-i\hbar c \:\partial \phi(x,t)/\partial x\;.
\end{eqnarray}
If $i\hbar \partial \phi/\partial t$ is negligible with respect to $2mc^2 \phi$ (with $c$
fixed at its actual value, not set equal to infinity) then $\phi$ can be replaced
by $-(i\hbar/2mc)\partial \Delta
\phi/\partial x$ in the second equation of (\ref{nonrel}) to obtain the Schr\"{o}dinger
equation $i\hbar \partial \Delta \phi(x,t)/\partial t = -(\hbar ^2/2m)\partial^2 \Delta
\phi(x,t)/\partial x^{\;2}$. If one further assumes that $|\Delta \phi(x,t)|^2 >>
|\phi(x,t)|^2$ and identifies $2^{-1/2} \Delta \phi(x,t)$ with the Schr\"{o}dinger
wave function $\psi_S(x,t)$ then the expressions (\ref{CB4}) and (\ref{CB4J}) immediately
lead to the desired nonrelativistic expressions, $\rho_S(x,t)=|\psi_S(x,t)|^2$ and
$J_S(x,t)=(\hbar/m)\Im[\psi_S^{*}(x,t)\partial\psi_S(x,t)/\partial x]$ respectively,
for the nonrelativistic probability and probability current densities.
Consistent with these considerations is the following simple choice of
initial $(t=0)$ wave function $\Psi(x_A,0)$ for the nonrelativistic regime: 
$\Psi_+(x_A,0)=-g\Psi_0(x_A)$ and $\Psi_-(x_A,0)=(1-g^2)^{1/2}\Psi_0(x_A)$
where $g$ is a real constant very close to $2^{-1/2}$ (see below)  and
$\Psi_0(x)=(2\pi)^{-1/4}(\Delta x)^{-1/2}\exp[-(2\Delta x)^{-2}(x-x_0)^2 +ik_0 x]$
is a minimum-uncertainty-product gaussian with initial centroid
$x_0$, initial variance $\Delta x$, mean wave vector $k_0$ and variance $\Delta k = 1/2\Delta x$.
In the numerical calculations presented below the constant $g$ is chosen so that the
characteristic velocity $v(x_A,0)\equiv J(x_A,0)/\rho(x_A,0)=(2g^2 - 1)c
\equiv v$ (independent of $x_A$) is equal to $v_0\equiv\hbar k_0/m$ with $v_0 << c$.

Now, from (\ref{KRR}), (\ref{KLL}) and (\ref{++1}) it immediately follows that
\begin{equation}
K_{++}^{(1)}(B;A)/K_{++/--}(B;A)\:=\:2v_{BA}/(c\pm v_{BA})\:\approx\:2v_{BA}/c
\end{equation}
and from (\ref{KRL}) and (\ref{+-1}) it follows that
\begin{eqnarray}
\frac{K_{+-}^{(1)}(B;A)}{K_{+-/-+}(B;A)}\:&=&\:\frac{J_0(l_{BA}/\lambda_c)
+[(c-v_{BA})/(c+v_{BA})]J_2(l_{BA}/\lambda_c)}{J_0(l_{BA}/\lambda_c)}\cr
\:&\approx&\:\frac{J_0(l_{BA}/\lambda_c)+J_2(l_{BA}/\lambda_c)-
(2v_{BA}/c)J_2(l_{BA}/\lambda_c)}{J_0(l_{BA}/\lambda_c)}.
\end{eqnarray}
In the regime $v_{BA}/c << 1$ under consideration,
$l_{BA}/\lambda_c\approx ct_{BA}/\lambda_c = (c/v_{BA})(x_{BA}/\lambda_c)
>> x_{BA}/\lambda_c >> 1$. Using the leading two terms in Hankel's
asymptotic expansions \cite{AbSteg} of $J_0(z)$ and $J_2(z)$ for large
argument $z$, i.e.
\begin{eqnarray}
J_0(z)\:&\approx&\:\biggl(\frac{2}{\pi z}\biggr)^{1/2}\biggl[ \cos \biggl(
z-\;\frac{\pi}{4}\biggr)\:+\:\frac{1}{8z}\sin \biggl( z-\frac{\pi}{4}\;\biggr)
\biggr]\;\;(|z|>>1) \cr
J_2(z)\:&\approx&\:\biggl(\frac{2}{\pi z}\biggr)^{1/2}\biggl[ \cos \biggl(
z-\frac{5\pi}{4}\biggr)-\frac{15}{8z}\sin \biggl( z-\frac{5 \pi}{4}
\biggr) \biggr]\:\:(|z|>>1)\;,
\end{eqnarray}
gives
\begin{equation}\label{ratio}
\frac{K_{+-}^{(1)}(B;A)}{K_{+-/-+}(B;A)}\;\approx\;2\;\biggl(\frac{v_{BA}}{c}
\biggr)\;
\frac{\cos \biggl(\frac{l_{BA}}{\lambda_c}\:-\:\frac{\pi}{4}\biggr)\:
+\:\frac{\lambda_c}{16x_{BA}}\:\sin\biggl(\frac{l_{BA}}{\lambda_c}\:-\:
\frac{\pi}{4}\biggr)}{\cos\biggl(\frac{l_{BA}}{\lambda_c}\:-\:\frac
{\pi}{4}\biggr)\:+\:\frac{\lambda_c}{8l_{BA}}\sin\biggl(\frac{l_{BA}}
{\lambda_c}\:-\:\frac{\pi}{4}\biggr)}\:.
\end{equation}
For $l_{BA}/\lambda_c = (4n+3)\pi/4$ with $n$ an integer, the right-hand-side
of (\ref{ratio}) is $1$. However, in a well-designed arrival-time
experiment for the wave function under discussion one would arrange
that $(X-x_0)>>\Delta x$ so that $\rho(x_A,0)$ is completely negligible for
$x_A\geq X$ and also that $\Delta x >> \lambda_c$ so that there is negligible
probability of generating particle-antiparticle pairs. Hence, $\lambda_c
/x_{BA}$  would be extremely small over the important range of $x_{BA}$.
Hence, the set of $x_{BA}$ values where the right-hand-side of (\ref{ratio})
is not close to $2v_{BA}/c$ is of small measure and can be ignored when
considering integrals over $x_A$, provided that $v_{BA}/c$ is not itself
extremely small (a rough estimate requires that $v_{BA}/c \;>>\;
\lambda_c/16\pi x_{BA}$).

\begin{figure}
{\includegraphics[width=3.35in]{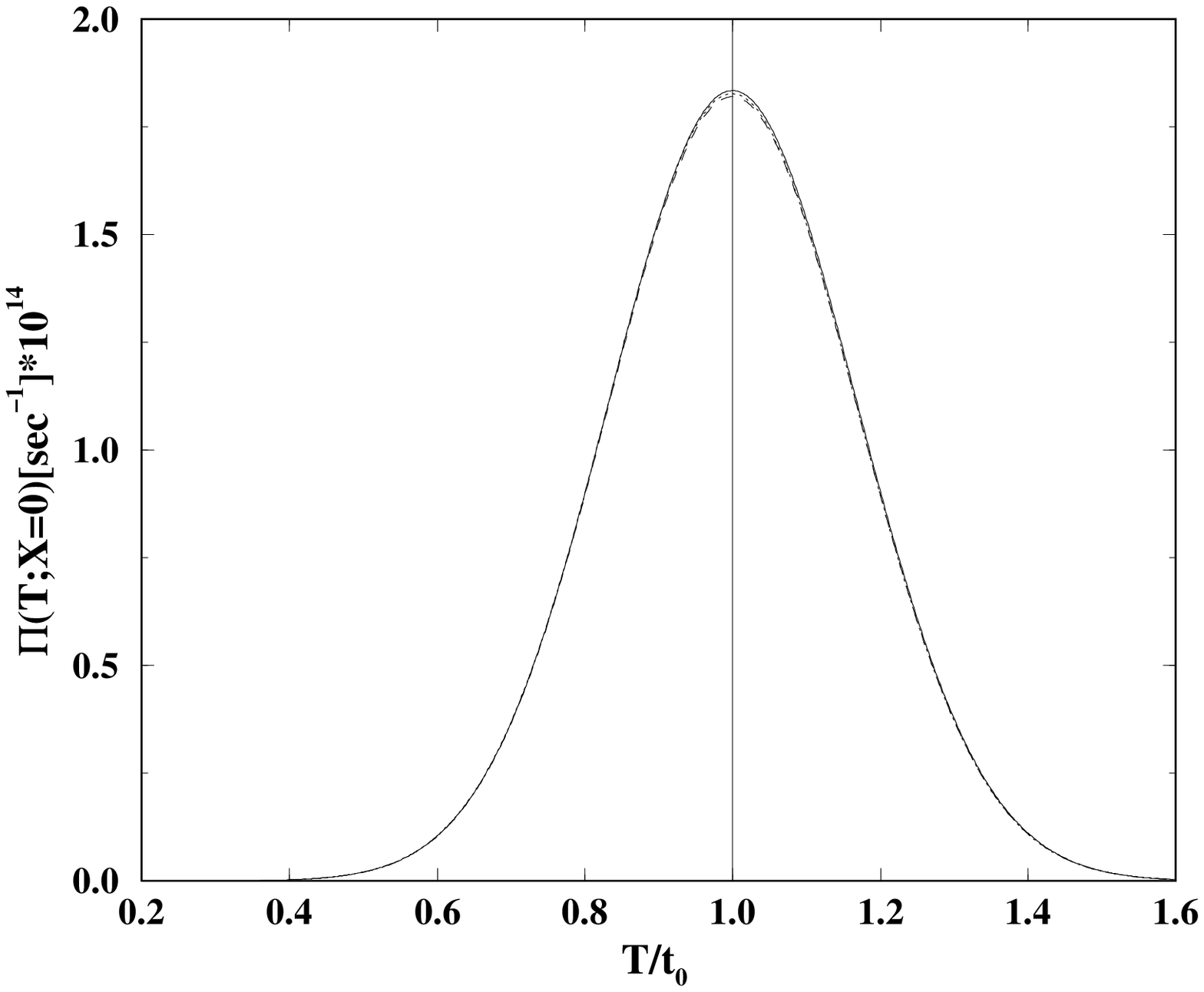}}
\caption[]{Arrival-time distribution $\Pi(T;0)$ (solid curve), scaled 
first-arrival contribution $\Pi^{(1)}(T;0)/2(v/c)^2$ (dotted line) 
and scaled interference contribution
$\Pi^{(1\times 2,3,...)}(T;0)/2(v/c)$ (dashed line) for the initial gaussian
wave function
described in the text. $k_0=1.00$ \AA$^{-1}$, $\Delta k = 0.02$ \AA$^{-1}$
and $x_0 = -6\Delta x$ with $\Delta x = 1/2\Delta k$; $v/c=v_0/c=3.862\times
10^{-3}$;
$t_0\:\equiv \: |x_0|/v_0\:=\:1.296\times 10^{-14}$ sec and $T_{max}=2t_0$.}
\label{ratio02}
\end{figure}

\begin{figure}
{\includegraphics[width=3.35in]{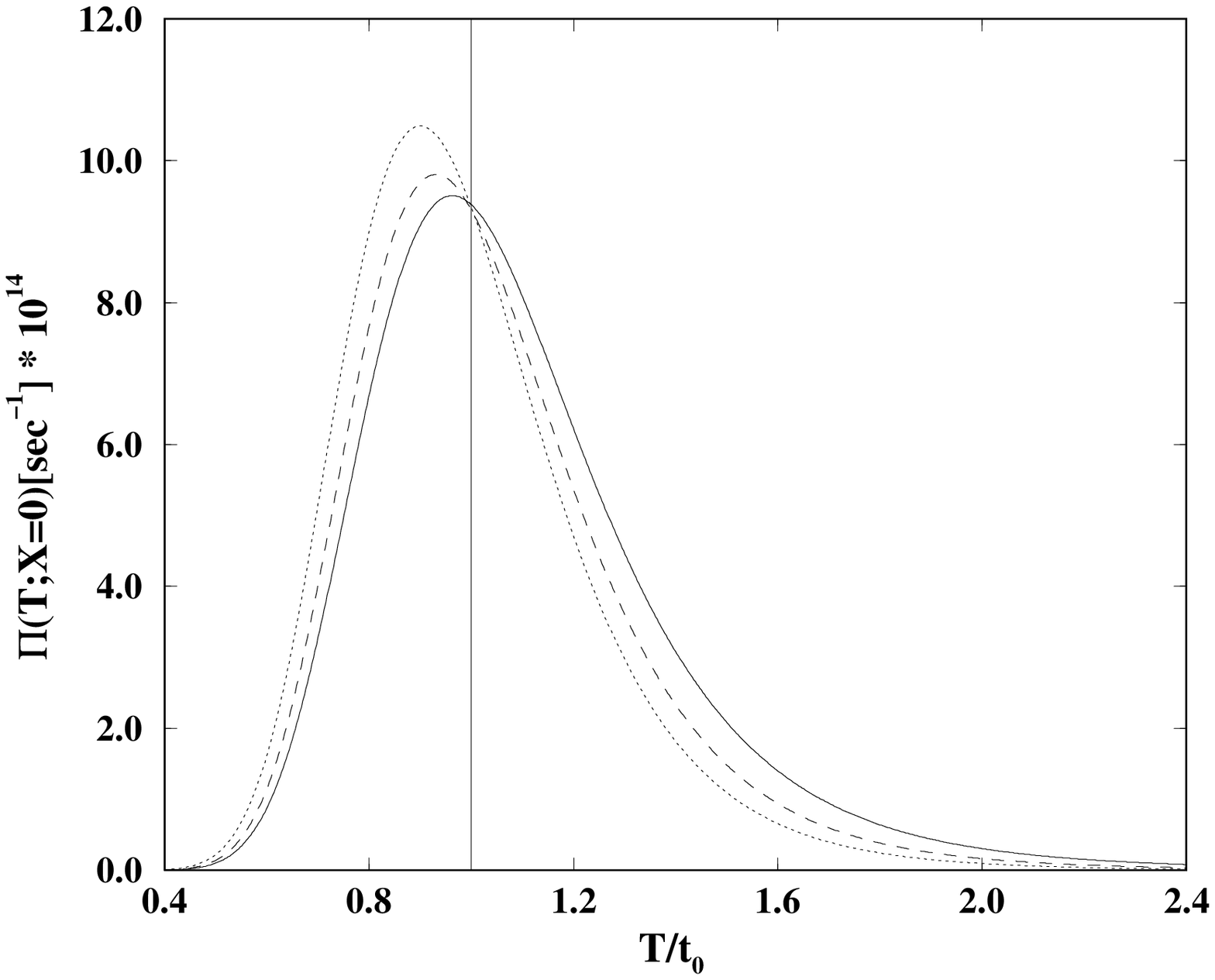}}
\caption[]{Arrival-time distribution $\Pi(T;0)$ (solid line), scaled
first-arrival
contribution $\Pi^{(1)}(T;0)/2(v/c)^2 $ (dotted line) and scaled interference
contribution
$\Pi^{(1\times 2,3,...)}(T,0)/2(v/c)$ (dashed line) for the initial
gaussian wave function
described in the text. $k_0=1.00$ \AA$^{-1}$, $\Delta k = 0.2$ \AA$^{-1}$
and $x_0 = -8\Delta x$ with $\Delta x = 1/2\Delta k$; $v/c=v_0/c=3.862\times
10^{-3}$;
$t_0=1.728\times 10^{-15}$ sec and $T_{max}=3t_0$.}
\label{ratio2}
\end{figure}

Taking into account that $-\Psi_-(x_A,0)\approx \Psi_+(x_A,0)$ for
$|v(x_A,0)|<<c$ and that the terms involving $K_{--}^{(1)}$ and
$K_{-+}^{(1)}$ are negligible when $|X-x_0|>>\Delta x$ then leads directly to
the estimates
\begin{equation}\label{est1}
\Pi^{(1)}(T;X)\;\approx\;(1/2)(2v/c)^2\;\Pi(T;X)\;=\;2(v/c)^2\;\Pi(T;X)\;,
\end{equation}
\begin{equation}\label{est2}
\Pi^{(1\times 2,3,...)}(T;X)\;\approx\;2(1/2)(2v/c)\;\Pi(T;X)\;=
\;2(v/c)\;\Pi(T;X)\;
\end{equation}
for the gaussian wave function under consideration. It should be noted
that $v_{BA}=(X-x_A)/T$ has been approximated by $v$ which is consistent
with $(X-x_0)>>\Delta x$ in the absence of significant wave packet spreading.
Fig. \ref{ratio02} and Fig. \ref{ratio2} show results
for $\Pi(T;0)$, $\Pi^{(1)}(T;0)/2(v/c)^2$ and $\Pi^{(1\times 2,3,...)}(T;0)
/2(v/c)$ obtained by numerical evaluation of (\ref{one}) to (\ref{cross})
for gaussian wave packets with $\Delta k /k_0 = 0.02$ and $0.2$
respectively. In the former case the above estimates are excellent
approximations; in the latter case, even though wave packet spreading is
more important, the estimates still provide good approximations for the very
large differences in overall scale between the three quantities.

\section{Concluding Remarks}

In summing up, it is interesting to make a qualitative comparison of the
results of
the Feynman path and Bohm trajectory approaches for investigating
arrival times for Dirac electrons.

In Bohmian mechanics \cite{BH,HOLLAND,CFG} an electron is postulated to be
an actually
existing point-like particle and an accompanying wave which guides its motion.
For a Dirac electron in the presence of a potential $V(\vec{r},t)$, the
time-evolution
of the guiding wave $\Psi(\vec{r},t)$ is described by the $3$D Dirac
equation and the
trajectory of the point-like particle is determined by the equation-of-motion
$d\vec{r}(t)/dt = [\vec{J}(\vec{r},t)/\rho(\vec{r},t)]|_{\vec{r}=\vec{r}(t)}$.
It is further postulated that, for an ensemble of electrons all prepared
in the same
initial state $\Psi(\vec{r},0)$, the probability of such a particle
{\it having}
initial position $\vec{r}^{\:(0)}\equiv \vec{r}(t=0)$ is given by
$\rho(\vec{r}^{\:(0)},0)$.
The various properties stated below for the intrinsic arrival times of the
point-like
particles of Bohm's theory follow readily from the fact that, for a given
initial wave
function, trajectories with different starting points $\vec{r}^{\:(0)}$
never
intersect or even touch each other \cite{TQM}.

Now, the expression for the intrinsic $1$D arrival-time distribution obtained
with either
approach can be cast in the same form as in classical mechanics, namely
\begin{equation}
\Pi(T;X)=\Pi_+(T;X)+\Pi_-(T;X)\;\; ; \;\; \Pi_{\pm}(T;X)\equiv \pm
J_{\pm}(X,T) \bigg /\int_0^{T_{max}} dt [J_+(X,t)-J_-(X,t)]\; ,
\end{equation}
where $J_+$ and $J_-$, respectively, are the right-going and left-going
components of
the probability current density $J$. The decomposition
$J\:=\:J_+\;+\;J_-$ is not uniquely
defined.  The decomposition associated with the fictitious particles
of the checkerboard
path approach, which move only at the speed of light $c$, is $J_{\pm}=\pm\:c\:
|\Psi_{\pm}|^2$ while that associated with the (assumed) actual particles
of the Bohm
trajectory approach, each of which moves at a variable speed that cannot
exceed $c$, is $J_{\pm}\:
=\:J \Theta[\pm J]$ where $\Theta$ is the unit step-function.
The former decomposition leads to an arrival-time distribution
$\Pi(T;X)$ proportional to the probability density $\rho(X,T)$ while
the latter leads to
one proportional to the absolute value of the probability current density,
i.e $|J(X,T)|$.
Moreover, unless one or other (or both) of the two components of $\Psi(X,t)$
is zero for
$T_1\leq t \leq  T_2 $, there are more arrivals --
many more if $T_2 -T_1$ is much larger than the Jacobson-Schulman correlation
time -- of
the fictitious particles at $X$ during that time interval than there are of
the supposed
actual particles of Bohm's theory. 

Given the probability current density $J(x,t)$ and using the non-crossing
property of Bohm
trajectories it is straightforward to decompose the intrinsic arrival-time
distribution
into contributions from first arrivals, from second arrivals, etc. with no
interference terms between different orders of arrival \cite{ROSS}.
In marked contrast to this, the decomposition based on Feynman 
checkerboard paths in general contains a nonzero interference term
between first and later (i.e., second, third, ...) 
arrivals so that from the calculation one cannot extract a well-defined
intrinsic first-arrival-time
distribution. In the nonrelativistic regime this interference term can be
very large
compared to the first-arrival term. Because of this and the extremely
small correlation length for reversal of direction ($\approx \lambda_c$
which is only about $10^{-3}$ of the diameter of an atom!), 
suppression of the interference term by decoherence \cite {JJH} 
within a time interval much less than $\lambda_c/c$ in duration
immediately following the instant of first arrival
would be very difficult, if not impossible, in a practical arrival-time
measurement. Unless this can be achieved, assuming that the first-arrival
times of the fictitious particles of the checkerboard model are directly
relevant to the arrival times measured in a time-of-flight experiment
on actual electrons is not justified. 

\acknowledgments{Support has been provided by Gobierno de 
Canarias (PI2000/111), Ministerio de Ciencia y Tecnolog\'\i a (BFM2001-3349
y BFM2000-0816-C03-03)
and the Basque Government (PI-1999-28).}


\begin{thebibliography}{99}

%
\bibitem{rev1} J. G. Muga, R. Sala and J. P. Palao, Superlattices and
Microstructures {\bf 23}, 833 (1998).
%
\bibitem{rev2} J. G. Muga and C. R. Leavens, Phys. Rep. {\bf 338}, 353 (2000).

%
\bibitem{MSE02} {\it Time in Quantum Mechanics},
edited by J. G. Muga, R. Sala Mayato and I. L. Egusquiza 
(Springer-Verlag, Berlin, 2002).
%
\bibitem{feynman}
R. P. Feynman and A. R. Hibbs, {\it Quantum Mechanics and Path Integrals}
(McGraw-Hill, New York, 1965), p.34-36.
%
\bibitem{yam1} N. Yamada and S. Takagi, Prog. Theor. Phys. {\bf 85}, 985
(1991).
%
\bibitem{yam2} N. Yamada and S. Takagi, Prog. Theor. Phys. {\bf 86}, 599
(1991).
%
\bibitem{JS}
T. Jacobson and L. S. Schulman, J. Phys. A {\bf17}, 375 (1984).
%
\bibitem{crl} C. R. Leavens, Phys. Lett. A {\bf 272}, 160 (2000).
%
\bibitem{Wigner}
E. P. Wigner, in {\it Aspects of Quantum Theory}, edited by A. Salam and
E. P. Wigner (Cambridge University Press, London, 1972).
%
\bibitem{GJKS}
B. Gaveau, T. Jacobson, M. Kac and L. S. Schulman, Phys. Rev. Lett. 
{\bf 53}, 419 (1984).
%
\bibitem{SPSAG}
L. S. Schulman, in {\it Path Summation: Achievements and Goals}, edited
by S. Lundqvist, A. Ranfagni, V. Sa-yakanit and L. S. Schulman (World
Scientific, Singapore, 1988).
%
\bibitem{kra}
C. Krattenthaler, {\it The enumeration of lattice paths with respect to their
number of turns}, in ``Advances in Combinatorial Methods and Applications to
Probability and Statistics'', pp. 29-58, N. Balakrishnan, Brirkh\"auser,
Boston (1997).
%
\bibitem{AbSteg}
M. Abramowitz and I. E. Stegum, {\it Handbook of Mathematical Functions}
(National Bureau of Standards, Washington, 1970).
%
\bibitem{BH}
D. Bohm and B. Hiley, {\it The Undivided Universe: An Ontological
Interpretation of Quantum Mechanics} (Routledge, London, 1993).
%
\bibitem{HOLLAND}
P. R. Holland, {\it The Quantum Theory of Motion} (Cambridge University Press,
Cambridge, 1993).
%
\bibitem{CFG}
J. T. Cushing, A. Fine and S. Goldstein (eds), {\it Bohmian Mechanics and
Quantum Theory: An Appraisal} (Kluwer, Dordrecht, 1996).
%
\bibitem{TQM}
C. R. Leavens, in {\it Time in Quantum Mechanics}, edited by J. G. Muga, 
R. Sala Mayato and I. L. Egusquiza (Springer, Berlin, 2002), Ch. 5.  
%
\bibitem{ROSS}
W. R. McKinnon and C. R. Leavens, Phys. Rev. A {\bf51}, 2748 (1995); 
X. Oriols, F. Mart\'{\i}n and J. Su\~n\'{e}, Phys. Rev. A {\bf54}, 2594 (1996).
%
\bibitem{JJH}
J. J. Halliwell, in {\it Time in Quantum Mechanics}, edited by J. G. Muga,
R. Sala Mayato and I. L. Egusquiza (Springer, Berlin, 2002), Ch. 6.  
%
\end{thebibliography}
\end{document}